\documentclass[
frontmatterverbose, 
 amsmath,amssymb,
 aps,
floatfix]{revtex4-1}
\usepackage{epsfig,graphics,amssymb,amsmath,graphicx,pifont,color,soul,esint,float}

\newcommand{\bl}[1]{\boldsymbol{#1}}
\newcommand{\dd}{\mathrm{d}}
\newcommand{\iinf}{\int_{-\infty}^{\infty}}
\providecommand{\BibitemShut}[1]{}

\begin{document}
\preprint{APS/123-QED}

\title{Description of clustering of inertial particles in turbulent flows via finite-time Lyapunov exponents}

\author{Mahdi Esmaily-Moghadam}
\author{Ali Mani}
\affiliation{Center for Turbulence Research, Stanford University}
  
\date{\today} 
\begin{abstract}

An asymptotic solution is derived for the motion of inertial particles exposed to Stokes drag in an unsteady random flow. 
This solution provides the finite-time Lyapunov exponents as a function of Stokes number and Lagrangian strain- and rotation-rates autocovariances.
The sum of these exponents, which corresponds to a concentration-weighted divergence of particle velocity field, is considered as a measure of clustering. 
For inertial particles dispersed in an isotropic turbulent flow our analysis predicts maximum clustering at an intermediate Stokes number and minimal clustering at small and large Stokes numbers. 
Direct numerical simulations are performed for quantitative validation of our analysis, showing a reasonable agreement between the two.
\end{abstract}
\maketitle

\section{Introduction} \label{intro}
Inertial particles in interaction with a spatially varying flow form regions of higher concentration.
This is known as particle clustering \cite{bec2003fractal}, segregation \cite{calzavarini2008quantifying}, or preferential concentration \cite{reade2000effect}.
The prevalence of particle-laden flow in areas of physics and engineering has inspired decades of research with applications ranging from planet and cloud formation to combustion and pharmaceutical applications \cite{eaton1994preferential,balachandar2010turbulent}.

These studies have established that clustering occurs when particles, characterized by their Stokes number, have a response time of order unity.
Much larger particles follow a ballistic trajectory that is independent of the flow.
Much smaller particles follow the fluid as tracers with minimal relative motion that is essential to formation of clusters.

Under a particular set of conditions, such as low particle Reynolds and Knudsen numbers, high particle-to-fluid density ratio, absence of body force, negligible finite size effects, and low mass loading ratio, particle motion is well described in these systems by the Stokes equation
\begin{equation}
\ddot X_i = \tau^{-1} \left( u_i - \dot X_i \right),
\label{stokes}
\end{equation}
in which $\bl X(t)$ is the position of a particle, $t$ is time, $\dot {(\bullet)} \equiv {\rm d}(\bullet)/ {\rm d} t$, $\bl u(\bl X, t)$ is fluid velocity, and $\tau$ is the particle relaxation time.
Although Eq. \eqref{stokes} seems a simple ordinary differential equation, it can exhibit a nonlinear chaotic behavior due to the dependence of its source term on $\bl X$ \cite{aref1986chaotic}.
This nonlinear behavior is central to the cluster formation process and even without all the assumptions associated with Eq. \eqref{stokes}, clustering follows a similar trend.
Therefore, we consider Eq. \eqref{stokes} as the starting point to probe the underlying physics of clustering of inertial particles at a more fundamental level.

Particle clustering can be characterized using multiple indices \cite{monchaux2012analyzing}.
One approach for characterization of particle clustering is based on the rate at which nearby particles separate or approach each other.
In other words, provided two particles at an initial infinitesimal distance $\|\bl r(0)\|$, they converge or diverge exponentially as
\begin{equation}
\| \bl r(t) \| = \|\bl r(0)\| e^{\lambda t}.
\label{LE_def}
\end{equation} 
$\lambda$ is called the finite-time Lyapunov exponent (FTLE) and converges to the Lyapunov exponent in an ergodic system as $t\to \infty$ \cite{shadden2005definition}.
As discussed in details in the following section, there are three exponents ($\lambda_1$, $\lambda_2$, and $\lambda_3$) in a 3-dimensional flow that determine the rate of expansion and contraction of a cloud of particles in three orthogonal directions.
These rate of expansion or contraction are related to the rate of change of the cloud volume, hence providing a measure of clustering.

Another approach for characterization of clustering is based on the statistics of particle concentration field, which is experimentally measurable and less mathematically abstract.
It is known that at the limit of small Stokes number, this concentration field is a fractal \cite{bec2003fractal}.
This allows establishing certain relationships between its different order statistics \cite{hentschel1983infinite}.
For instance, the scaling exponent, $\zeta(n)$, can be obtained from $m^n = D^\zeta$, in which $m^n(D)$ is the average mass of particles inside a infinitesimal sphere with diameter $D$.
This index produces a range of values depending on the level of clustering, as the clustering of randomly distributed particles toward a 2D manifold reduces $\zeta$ from $3n$ to $2n$ in a 3D flow.
More interestingly, for a fractal concentration field $\zeta$ can be directly relate to the Lyapunov exponents, i.e. the first aforementioned index \cite{bec2004multifractal,wilkinson2010correlation}.
Therefore, the FTLEs also provide estimate for a more experimentally-accessible clustering index and hence are of prime importance for characterization of particle clustering. 

Various approaches have been adopted to relate these indices to the background flow and Stokes number. 
Experimental observation \cite{fessler1994preferential,aliseda2002effect,salazar2008experimental,saw2008inertial} and direct numerical simulations (DNS) \cite{ray2011preferential,calzavarini2008quantifying,tagawa2012three,goto2008sweep} have shown that Stokes number of order unity based on the Kolmogorov time scale is associated with the strongest regime of clustering. 
Analytically, by taking the divergence of Eq. \eqref{stokes} under the assumptions $\ddot {\bl X} \approx \mathrm{D} \bl u/\mathrm{D}t$ and $\nabla \cdot \bl u = 0$, it can be shown \cite{robinson1956motion,maxey1987gravitational}
\begin{equation}
-\nabla \cdot \dot {\bl X} \approx \tau \nabla \cdot \left(\bl u \cdot \nabla \bl u \right) = \tau(\|\bl S\|^2 - \|\bl \Omega\|^2),
\label{maxyes}
\end{equation}
in which $\bl S$ and $\bl \Omega$ are the fluid strain- and rotation-rate tensors, respectively.
Since this divergence is equal to the sum of FTLEs (i.e. $\lambda_1+\lambda_2+\lambda_3$), Eq. \eqref{maxyes} directly relates both indices described above to the background flow. 
Having the second invariant of velocity gradient tensor (subtracted from the first invariant) on the right hand side of this expression explains that the most hospitable zone for clustering are those with high strain and low rotation rate.
Although this relation provides information about both indices through the sum of FTLEs, it is linearly proportional to $\tau$ and fails to predict the non-monothonic behavior of particle clustering versus Stokes number. 

The goal of this study is finding a correction to Eq. \eqref{maxyes} that not only reproduces it at the limit of small Stokes number (the regime that Eq. \eqref{maxyes} is applicable to), but also is applicable to a wider range of Stokes numbers.
With this motivation in mind, in what follows, an asymptotic solution is derived for Eq. \eqref{stokes} that quantifies FTLEs associated with inertial particle pairs.
This solution expresses the sum of FTLEs as a function of Stokes number and background flow statistics, hence can be related to both indices described above.
Our analysis considers a general representation of flow with oscillatory modes of strain and rotation acting simultaneously over all possible frequencies.
We show how the clustering problem turns into an eigenvalue problem with its eigenvalues and eigenvectors representing magnitude and directions of contractions, respectively.
We also show how a particle, depending on its Stokes number, may filter or resonate with different frequencies. 
A validation of our analysis is also presented through a quantitative comparison against DNS of particle laden homogeneous isotropic turbulent (HIT) flow.

\section{Analytical derivation}
\begin{figure}
\begin{center}
\includegraphics[width=0.4\textwidth]{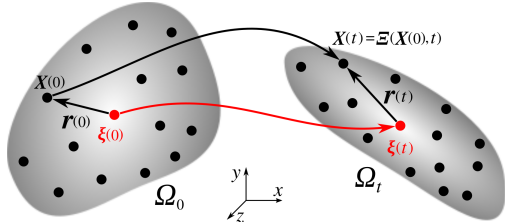}
\caption{Schematic of a cloud of particles, $\Omega_t$, defined as a 3D manifold occupied by a collection of particles.
This cloud, which undergoes deformation characterized by $\bl \Xi$, may expand or contract toward a central point $\bl \xi$.}
\label{fig:particle_cloud}
\end{center}
\end{figure}

As the starting point, we consider a volume occupied by a collection of nearby particles, which we call a cloud.
This cloud is initially denoted by $\Omega_0$ and evolves over time with $\bl \Xi: \Omega_0 \to \Omega_t$ (Fig. \ref{fig:particle_cloud}).
By taking function $\bl \Xi$ to satisfy Eq. \eqref{stokes}, $\Omega_t = \{\bl X(t) \mid \bl X(t) = \bl \Xi(\bl X(0),t), \bl X(0) \in \Omega_0$\}. 
By following $\Omega_t$, the problem is formulated on a reference frame that moves with the cloud. 

Denoting the central point of the cloud by $\bl \xi$, relative motion of a particle located at $\bl X \in \Omega_t$ to the central point is $\bl r(t) = \bl X(t) - \bl \xi(t)$ (Fig. \ref{fig:particle_cloud}). 
Substituting this relation in Eq. \eqref{stokes} gives
\begin{equation}
\tau (\ddot \xi_i + \ddot r_i) + \dot \xi_i + \dot r_i = u_i(\bl \xi + \bl r,t).
\label{stokes_relative}
\end{equation}
By bounding the size of the cloud to be less than the Kolmogorov length scale $\eta$, we ensure that the cloud is not in the inertial interval and probes a smooth velocity field.
Hence, from the Taylor series expansion
\begin{equation}
u_i(\bl \xi + \bl r,t) = u_i(\bl \xi,t) + u_{i,j}(\bl \xi,t) r_j + \mathcal O\left(\| \bl r\|^2 \right), 
\label{tylor_expansion}
\end{equation}
in which $(\bullet)_{,i} \equiv \partial(\bullet)/\partial x_i$.  
Neglecting the higher order terms in Eq. \eqref{tylor_expansion} and using $\tau \ddot \xi_i + \dot \xi_i = u_i(\bl \xi,t)$ to simplify Eq. \eqref{stokes_relative} leads to
\begin{equation}
\tau \ddot r_i + \dot r_i = u_{i,j}(\bl \xi,t) r_j,
\label{reduced_stokes}
\end{equation}
which is an exact linearized form of Eq. \eqref{stokes}, provided $\|\bl r\|$ is sufficiently small.
Note that, we have derived Eq. \eqref{reduced_stokes} from Eq. \eqref{stokes} in several steps to build physical intuition.
This equation can be obtained by taking the derivative of Eq. \eqref{stokes} with respect to $\bl X(0)$ \cite{ijzermans2010segregation}.

Fluid velocity is expressed in a continuous Lagrangian Fourier space that follows the cloud, 
\begin{equation}
u_i(\bl \xi, t) = \sum_\omega \tilde u_i(\omega) e^{\hat i \omega t},
\label{u_fft}
\end{equation}
in which $\omega = (-\infty, \infty)$.
Since finding $\bl r$ for an individual cloud is of interest, dependence of $\tilde {\bl u}$ on $\bl \xi$ is dropped from Eq. \eqref{u_fft}.
From Eqs. \eqref{u_fft} and \eqref{reduced_stokes}
\begin{equation}
\tau\ddot r_i + \dot r_i = \sum_\omega \tilde u_{i,j}(\omega) r_j e^{ \hat i \omega t }.
\label{governing_eq}
\end{equation}
These three ordinary differential equations relate particle motion to the harmonics of the velocity gradient tensor sampled at the cloud center.
One significance of this relation is that the fluid velocity gradient is characterized by the Kolmogorov scale, and thus explains our earlier use of Kolmogorov units.
This, however, does not imply that Eq. \eqref{governing_eq} only captures a subset of existing scales of the flow. 
Since $u_{i,j}$ is a general function of time, the full turbulent spectrum is accounted for in our analysis, including the longer time scales of the inertial range. 
Additionally, due to the turbulence intermittency $\tilde u_{i,j}$ can not be represented as a unique set of harmonic functions. 
However, it is still possible to derive a generic solution for the clustering index, as shown later in this section. 

Denoting the volume of the cloud $\Omega_t$ by $V(\Omega_t)$, we define a clustering index, $\mathcal C$, such that 
\begin{equation}
V(\Omega_t) = V(\Omega_0) \exp(-\mathcal C t).
\label{C_def}
\end{equation}
With this definition the cloud might rotate or even expand in a particular direction, but it clusters as long as the number density within the cloud increases. 
To quantify $\mathcal C$, we search for an asymptotic solution to Eq. \eqref{governing_eq} with a form of 
\begin{equation}
r_i = e^{\lambda t} \sum_\omega A_i(\omega) e^{\hat i \omega t},
\label{sol_guess}
\end{equation}

in which $\lambda$ and $A_i$ are the free eigenvalues and eigenfunctions to be determined.
Considering the upper bound on $\|\bl r\|$, $\lambda$ is the FTLE associated with particle pairs at $\bl \xi$ and $\bl \xi + \bl r$.
The solution form in Eq. \eqref{sol_guess} allows for oscillation over all possible frequencies, $\omega$, as well as contraction or expansion characterized by $\lambda$.
Specifically, we seek contractions and expansions that persist over time scales longer than particle relaxation time and thus regimes with 
\begin{equation}
|\lambda\tau| \ll 1,
\label{lambda_assumption}
\end{equation}
is of particular interest. 
As detailed in the Appendix, this equation can be expanded at $|\omega| < |\lambda|$ and $|\omega| > |\lambda|$ and simplified using Eq. \eqref{lambda_assumption} to obtain a 3$\times$3 eigenvalue problem as
\begin{equation}
\bl \psi \bl A^0 = \lambda \bl A^0,
\label{eigen_value_problem}
\end{equation}
in which $\bl A^0$ is the displacement vector associated with the low frequency oscillations, i.e. $|\omega| < |\lambda|$, and 
\begin{equation}
\psi_{ij} \equiv -\sum_\omega \frac{\tau}{ 1 + (\tau \omega)^2 } \tilde u_{i,k}^* \tilde u_{k,j},
\label{psi}
\end{equation}
in which $\tilde {\bl u}^*$ is the complex conjugate of $\tilde {\bl u}$.
The three eigenvectors of $\bl \psi$ represent the principal directions at which the cloud experiences pure contraction or expansion. 
The eigenvalues associated with each direction represent the rate of contraction (real$(\lambda) < 0$) or expansion (real$(\lambda) > 0$). 
Numerical investigation shows that generally one of the eigenvalues has a positive and one has a negative real part, hence most clouds expand and contract at the same time, which is consistent with previous reports \cite{bec2005multifractal}.
The imaginary part of each eigenvalue accounts for the mean rotation.

While the instantaneous rate of expansion or contraction of a cloud is subject to oscillations, the long term change of $V(\Omega_t)$ is solely controlled by $\lambda$ and can be computed from the product of contraction or expansion factors in the principal directions as
\begin{equation}
V(\Omega_t) = V(\Omega_0) \exp(\Lambda_{ii} t).
\label{v_omega} 
\end{equation}
In this relation, $\bl \Lambda$ is the eigenvalue matrix of $\bl \psi$, hence $\Lambda_{ii}$ is the sum of three FTLEs, $\lambda$'s.
Considering our earlier definition of clustering index, Eq. \eqref{v_omega} provides $\Lambda_{ii}$ as the measure of clustering index for a single cloud of particles. 
Given invariance of trace, $\Lambda_{ii}$ can be directly computed from $\psi_{ii}$.
Since individual clouds cluster differently in a turbulent flow, it is necessary to average over several clouds to obtain an ensemble averaged clustering index.
Hence from Eqs. \eqref{C_def} and \eqref{v_omega}
\begin{equation}
\mathcal C \approx -\langle \Lambda_{ii} \rangle  = -\langle \psi_{ii} \rangle = \sum_\omega \frac{\tau}{1 + (\tau \omega)^2} \langle \tilde u_{i,j}^* \tilde u_{j,i} \rangle,
\label{c_index}
\end{equation}
in which $\langle \bullet \rangle$ denotes ensemble averaging and $\langle \tilde u_{i,j}^* \tilde u_{j,i} \rangle = \langle \tilde{\bl S}^* \tilde{\bl S} \rangle - \langle \tilde {\bl \Omega}^* \tilde {\bl \Omega} \rangle$.
Using the convolution theorem, Eq. \eqref{c_index} can be expressed in terms of a continuous integral as
\begin{equation}
\mathcal C(\tau) = \iinf \frac{\tau \phi(\omega)}{1 + (\tau \omega)^2} \dd\omega,
\label{clustering_index}
\end{equation}
in which  
\begin{equation}
\phi(\omega;\tau) \equiv \tilde {\mathcal R}_{\{\bl S\}}  - \tilde {\mathcal R}_{\{\bl \Omega\}}, 
\label{phi}
\end{equation}
where $\tilde {\mathcal R}$ represents the Fourier transform of the autocovariance functions defined as $\mathcal R_{\{\bl S\}}(t) \equiv \langle \bl S(t') \bl S(t'+t) \rangle$ and $\mathcal R_{\{\bl \Omega\}}(t) \equiv \langle \bl \Omega(t') \bl \Omega(t'+t) \rangle$.
Since $\bl S$ and $\bl \Omega$ are computed at the location of the cloud, $\phi$ is different from Eulerian averaged quantity and is thus dependent on $\tau$.

Interestingly in the limit of $\tau^* = \tau/\tau_{\eta} \ll 1$, in which $\tau_\eta$ is the Kolmogorov time scale of the flow, Eq. \eqref{clustering_index} reduces to 
\begin{equation}
\mathcal C = \tau \iinf \phi \dd\omega = \tau \left( \langle \bl S^2 \rangle - \langle \bl \Omega^2 \rangle \right),
\end{equation}
which is identical to Eq. \eqref{maxyes}. 
Note that Eq. \eqref{maxyes} is validated in the limit of $\tau^* < \mathcal O(1)$, showing its close connection with particle clustering \cite{wang1993settling,rani2003evaluation}.
Here using a completely different approach, we derived a more general expression for clustering that is applicable to a wider range of $\tau^*$'s.
Among all flow parameters, this relation only depends on the first and second invariants of the velocity gradient tensor, confirming the dominant role of viscous scales on producing fluctuation in the particle concentration field \cite{balkovsky2001intermittent}.
 
Equation \eqref{clustering_index} is the most important result of this analysis, which at the first glance predicts maximum particle clustering at intermediate $\tau$ when $\phi > 0$.
Additionally, neglecting the dependence of $\phi$ on $\tau$, it predicts the decay of clustering index for $\tau \to 0$ or $\tau \to \infty$ proportional and inversely proportional to $\tau$, respectively. 
Furthermore, the fact that the product of $\tau$ and $\omega$ appeared in the denominator of Eq. \eqref{clustering_index} explains the unresponsiveness of larger particles to the fast oscillations of small flow features. 

\section{Numerical validation}
\begin{figure}
\begin{center}
\includegraphics[width=0.5\textwidth]{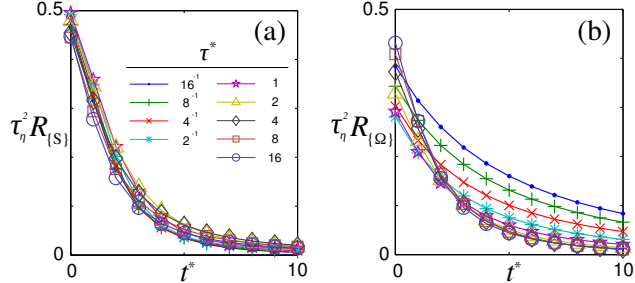}
\caption{The Lagrangian (a) strain- and (b) rotation-rate autocorrolation functions versus time ($t^*=t/\tau_{\eta}$) at different Stokes numbers ($\tau^* = \tau/\tau_\eta$).}
\label{fig:G_v_t} 
\end{center}
\end{figure}

In this section, we quantify the outcome of this analysis, i.e. Eq. \eqref{clustering_index}, in a HIT flow and compare it against direct numerical results.
DNS of a triply periodic incompressible flow at Re$_{\lambda}$=100 is performed using a $256^3$ numerical grid. 
Turbulence is maintained using a linear forcing term \cite{rosales2005linear}.
Eq. \eqref{stokes} is solved for particles with $\tau^*=2^p$, $p\in\{-4,\ldots,4\}$.
Simulations are initialized with $10^5$ randomly seeded particles at each Stokes number and continued for several large eddy turn-over time.
Starting from this time-evolved distribution, velocity gradient tensor is recorded at the position of each particle for 1200$\tau_{\eta}$ and autocovariance functions are computed at each $\tau^*$ accordingly (Fig. \ref{fig:G_v_t}).

At the limit of very large Stokes number, particles are not responsive to the velocity fluctuations and follow a trajectory that is uncorrelated with the flow.
As a result, the Lagrangian statistics at this limit converge to the Eulerian statistics, which are by definition obtained from a fixed location in space that is uncorrelated with the flow.
Additionally, the Eulerian strain- and rotation-rate autocorrelation functions are equal on a periodic domain.
Therefore, as confirmed by Fig. \ref{fig:G_v_t}, the Lagrangian strain- and rotation-rate autocorrelation functions converge to the same value at the limit of large Stokes number.
Moreover, since the strain-rate autocovariance function is fairly independent of the Stokes number, one may conclude that the trajectory of inertial particles is poorly correlated with the flow strain field.
This is not the case, however, with the flow rotation field.
The rotation-rate autocovariance function at small Stokes numbers is lower and higher than the corresponding curve at large Stokes number at shorter and longer time periods, respectively (Fig. \ref{fig:G_v_t}).
This indicates that for $\tau^*\le \mathcal O(1)$ particles are repelled from the fast toward the slower vortical regions of the flow.
Note that the time scale that this cross-over occurs increases as the Stokes number increases.
These trends are compatible with the literature  \cite{rani2003evaluation}. 

\begin{figure}
\begin{center}
\includegraphics[width=0.45\textwidth]{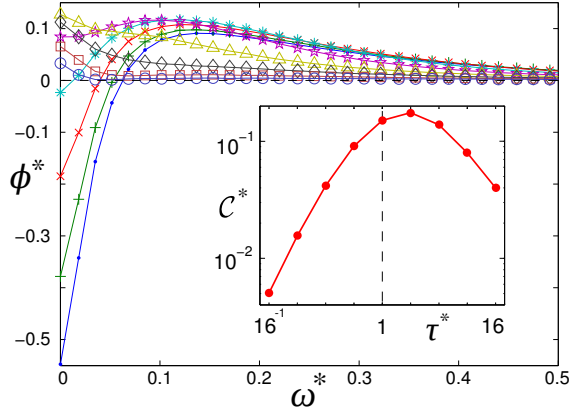}
\caption{$\phi^*=\tau_\eta \phi$ versus $\omega^*=\tau_{\eta}\omega/\pi$ at different Stokes numbers.
For symbols, see Fig. \ref{fig:G_v_t}.
Note $\int \phi^* \dd \omega^* \propto \sqrt{\tau^*}$ as $\tau^* \to 0$ and $\phi^*(\omega^* \ll 1) \propto 1/\tau^*$ as $\tau^* \to \infty$.
Error bars (not shown) are of $\mathcal O(10^{-4})$. 
Inset: Analytical prediction of the clustering index via Eq. \eqref{clustering_index} as a function of $\tau^*$.
Note maximum clustering occurs at $\tau^* = \mathcal O(1)$.}
\label{fig:phi} 
\end{center}
\end{figure}

Using Eq. \eqref{phi}, $\phi$ is computed and shown in Fig. \ref{fig:phi}.
A biased sampling of strain- and rotation-rate-rich regions of the flow by inertial particles produces a non-zero $\phi$.
This figure also confirms that particles with $\tau^*<1$ tend to follow slow vortical features ($\phi^*<0$ at $\omega^* \ll 1$ and $\tau^* < 1$), while particles with $\tau^* \gg 1$ experience strain- and rotation-rates equally ($\phi^* \to 0$ as $\tau^* \to \infty$).
Most notably $\phi^*$ is positive across all frequencies for $\tau^* \ge \mathcal O(1)$.

\begin{figure}
\begin{center}
\includegraphics[width=0.45\textwidth]{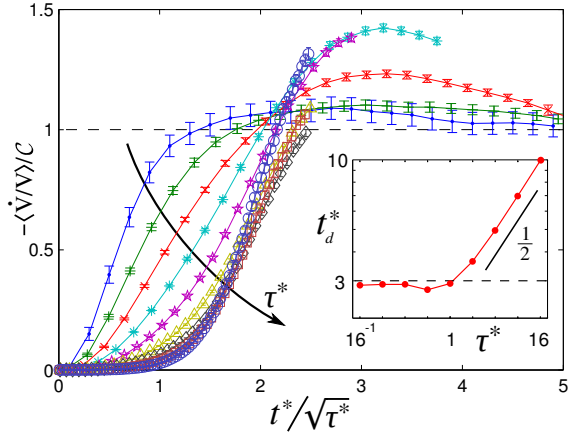}
\caption{Time-evolution of ensemble average of measured $-\dot V/V$ of clouds normalized by theoretically predicted $\mathcal C$ at different $\tau^*$.
For symbols, see Fig. \ref{fig:G_v_t}.
Note the collapse of curves with $\tau^*>1$ in the transient period and also curves with $\tau^*<1$ approaching unity for $t^*>1$.
Inset: Variation of distortion-time, $t_{\rm d}^*$, versus $\tau^*$.}
\label{fig:C} 
\end{center}
\end{figure}

$\mathcal C$ is calculated theoretically, using Eq. \eqref{clustering_index}, and shown in the inset of Fig. \ref{fig:C}. 
In these calculations, Eq. \eqref{lambda_assumption} is satisfied well with $|\lambda \tau|$ ranging from $10^{-4}$ at $\tau^*=16^{-1}$ to 0.21 at $\tau^* \ge 8$.
$\mathcal C^*$ has a non-monotonic behavior with $\max(\mathcal C^*) = \mathcal C^*(\tau^*=2) \approx 0.18$ on the discrete set of investigated $\tau^*$.
For $\tau^* \gg 1$, $\mathcal C^* \propto 1/\tau^*$, which is due to $\lim_{\tau^* \to \infty} \phi^*(\omega^* \ll 1) \propto 1/\tau^*$.
For $\tau^* \ll 1$, $\mathcal C^* \propto \tau^{*3/2}$, which is due to $\lim_{\tau^* \to 0}\iinf \phi^* \dd\omega^* \propto \sqrt {\tau^*}$.
This slope of 1.5 has also been reported to be 2 at this regime of Stokes number \cite{bec2003fractal,ijzermans2010segregation}. 
This difference can be ascribed to the relatively large value of the smallest Stokes number that is considered in this study and also the use of synthetic flows as oppose to HIT in the previous studies.

\begin{figure}
\begin{center}
\includegraphics[width=0.45\textwidth]{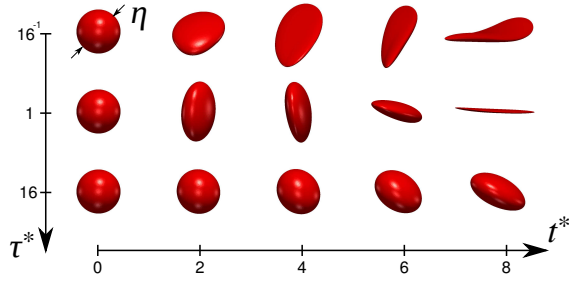}
\caption{Deformation of three arbitrary spherical clouds in a HIT flow (particles are not shown).
These clouds are constructed as the convex hull of a set of inertial particles seeded on a spherical shell. 
From top to bottom: $\tau^*=16^{-1}$, 1, and 16. 
While the cloud with $\tau^*=16^{-1}$ deforms significantly, its volume remains within 20\% of the initial volume.
$\tau^*=1$ cloud contracts significantly with a five-fold decrease in volume.
$\tau^*=16$ cloud remains almost spherical with a minimal 5\% change in volume.}
\label{fig:cloud_def} 
\end{center}
\end{figure}

To validate this analysis, a numerical procedure is devised to compute contraction rate of clouds of particles from the DNS.
A cloud is constructed by seeding randomly-distributed particles on a spherical shell with diameter equal to the Kolmogorov length.
A tetrahedral mesh is constructed by taking these particles as the vertices of the mesh.
The volume of the cloud is computed numerically by integrating the volume inside the mesh.
Fig. \ref{fig:cloud_def} shows time evolution of three arbitrary clouds constructed in this manner.

To obtain an ensemble averaged $\mathcal C(\tau)$ measured from the DNS, 25 simulations, each ran for $10\tau_\eta$ with $10^4$ clouds, are performed. 
In these calculations, the velocity of randomly seeded particles on the spherical shell is set to that of the sampled particle at the center.
Hence, $\mathcal C$ is initially zero and increases gradually and approaches to a fixed value once the transient period is passed (Fig. \ref{fig:C}).
This transient time is proportional to $\tau$ and $\sqrt \tau$ for clouds with $\tau^* \le \mathcal O(1)$ and $\tau^* \gg 1$, respectively.
This proportionality-exponent of transient time versus $\tau$ can be analytically extracted from Eq. \eqref{reduced_stokes} by taking $u_{i,j} = 1/\tau_{\eta}\delta_{ij}$, in which $\delta_{ij}$ is the Kronecker delta.

\begin{figure}
\begin{center}
\includegraphics[width=0.45\textwidth]{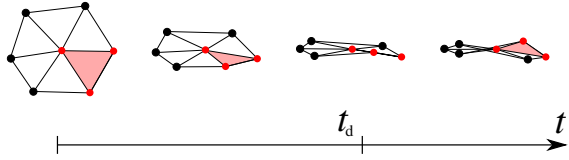}
\caption{Schematic of a 2D cloud, which consists of seven particles, deforming over time. 
Total area is calculated as the sum of six elements area. 
Distortion time, $t_{\rm d}$, corresponds to time that the area of at least one of the elements (distinguished by red) becomes negative.}
\label{fig:td} 
\end{center}
\end{figure}

After an extended period of time, clouds become \emph{distorted} due to the large deformation or surface cross-overs. 
This is shown schematically for a 2-dimensional cloud in Fig. \ref{fig:td}.
By definition, distortion occurs when the volume of an element of the mesh becomes negative.
To detect a negative volume, we monitor the sign of the determinant of the Jacobian associated with the element mapping. 
Once a sign change occurs, the cloud is excluded from our calculation.
Physically, distortion occurs because particles that are at the same position may have different velocities, which is a direct consequence of the compressibility of particle velocity field. 
The simulation is stopped once 5\% of all clouds become distorted. 
This is to avoid a bias toward expanding clouds. 
Our results show that this occurs at $t^*_d = t_{\rm d}/\tau_\eta \approx 3$ for $\tau^*<1$ and $t^*_d = 2.5\sqrt{\tau^*}$ for $\tau^*>1$ (inset of Fig. \ref{fig:C}). 

Figure \ref{fig:C} shows $\mathcal C$ measured from the DNS (i.e. $\langle -\dot V/V \rangle$) normalized by theoretically predicted $\mathcal C$ as a function of time.
Without this normalization, curves at the investigated range of $\tau^*$ would vary by orders of magnitude (inset of Fig. \ref{fig:phi}), while after normalization, they all reduce to a range of order one. 
Closer comparison between the DNS measurements and analytical results can be made at two limits of $\tau^* < 1$ and $\tau^* > 1$. 
For $\tau^* < 1$, corresponding curves in Fig. \ref{fig:C} asymptote to one, hence, $\mathcal C$ measured from the DNS is equal to the analytically-derived $\mathcal C$ at small Stokes numbers.
For $\tau^* > 1$, while long-term variation of $\mathcal C$ from the DNS is not available, our normalization still leads to $\mathcal O(1)$ quantities.
In addition, curves with $\tau^*>1$ collapse in the short-term transient period when normalized by analytical $\mathcal C$.
For synthetic flows, it has been shown that for $\tau^*\gg1$ and $t^*\gg1$, $\mathcal C$ becomes negative, indicating the dominance of expansion over contraction \cite{meneguz2011statistical,ijzermans2010segregation}. 
Therefore, long term behavior of particles with large Stokes number remains as a subject of future studies.

Our result analytically explains the outcome of previous investigations that have characterized clustering via other indices, yet consistently finding maximum clustering at $\tau^*=\mathcal O(1)$ \cite{wang1993settling,bec2007heavy,reade2000effect,ray2011preferential}.
The outcome of this study has direct implications in developing two-particle models for large-eddy simulation of particle-laden turbulent flows \cite{ray2013investigation}.
Additionally, the clustering index that was employed in this study is directly related to
\begin{equation}
d_{\rm L} = 3 - (\lambda_1+\lambda_2+\lambda_3)/\lambda_3 = 3 + \mathcal C/\lambda_3,
\label{dl}
\end{equation}
which is the Lyapunov dimension introduced by Kaplan and York \cite{kaplan1979functional}.
In this equation, which is valid for a 3-dimensional flow, exponents are sorted in a descending order such that $\lambda_3<\lambda_2<\lambda_1$. 
In practical cases $\lambda_3<0$, hence for $\mathcal C>0$, $d_{\rm L}$ is smaller than the number of spatial dimensions. 
$d_{\rm L}$ can be interpreted as a space dimension that is filled by the particle fractals.
By estimating $\lambda_3$, one may employ the result of this analysis to estimate the dimension of particle fractals.
As it was discussed in Section \ref{intro}, this dimension can be also related to the particle concentration field statistics through $\zeta$.

\section{Conclusion}
In summary, we derived an asymptotic solution for clustering of inertial particles, defined as the summation of the FTLEs in three principal directions, in regimes relevant to particle-laden turbulent flows.
The analytical model that we derived shows clustering is maximized at $\tau^* = \mathcal O(1)$ and decays as $\tau^* \to 0$ or $\infty$, which qualitatively is consistent with previous measurements.
This non-monotonic variation of clustering versus Stokes number was predicted without employing any tunable parameters.
The developed model exactly reproduces previously established results in the limit of small Stokes number \cite{maxey1987gravitational,rani2003evaluation}.
A remarkable agreement was observed between our analysis and DNS in HIT flow.
Additionally the developed framework reveals the contribution of different time scales in turbulence to the clustering phenomenon via the defined $\phi(\omega)$.
The presented analysis is generic and can be potentially applied to compressible and anisotropic turbulent flows, hence warranting future studies with a wider range of scenarios.

\begin{acknowledgments}
We thank Professors Parviz Moin and John Eaton for fruitful discussions. 
This work was supported by the United States Department of Energy under the Predictive Science Academic Alliance Program 2 (PSAAP2) at Stanford University.
\end{acknowledgments}

\appendix*
\section{Derivation of Eq. \eqref{eigen_value_problem}}

Step by step derivation of Eq. \eqref{eigen_value_problem} is included in this appendix. 
In the following, we use short hand notations $\bl A^m \equiv \bl A(\omega^m)$ and $\tilde {\bl u}^{m-n} \equiv \tilde {\bl u}(\omega^m - \omega^n)$. 

Substitute Eq. \eqref{sol_guess} in \eqref{governing_eq} yields
\begin{equation}
\sum_{\omega^m} \left( \tau (\hat i \omega^m + \lambda )^2 + \hat i \omega^m + \lambda \right) A_i^m e^{\hat i \omega^m t} = \sum_{\omega^l} \sum_{\omega^n}\tilde u_{i,j}^l A_j^n e^{ \hat i ( \omega^l + \omega^n ) t }.
\label{sol_expanded_form}
\end{equation}

Since summations are calculated over infinite intervals, we take $\omega^l = \omega^m - \omega^n$ to simplify Eq. \eqref{sol_expanded_form} to 
\begin{equation}
\sum_{\omega^m} \left( \tau (\hat i \omega^m + \lambda )^2 + \hat i \omega^m + \lambda \right) A_i^m e^{\hat i \omega^m t} = \sum_{\omega^m} \sum_{\omega^n} \tilde u_{i,j}^{m - n} A_j^n e^{ \hat i \omega^m t },
\label{sol_expanded_int}
\end{equation}
which must hold at any $t$. 
This is achieved by ensuring
\begin{equation}
\left( \tau (\hat i \omega^m + \lambda )^2 + \hat i \omega^m + \lambda \right) A_i^m = \sum_{\omega^n} \tilde u_{i,j}^{m - n} A_j^n,
\label{sol_expanded_simp}
\end{equation}
is satisfied for any $\omega^m$. 
In the following, we enforce this condition separately for small and large $\omega^m$.

Approximating Eq. \eqref{sol_expanded_simp} for $|\omega^m| < |\lambda|$ yields
\begin{equation}
\left(\tau \lambda + 1\right)\lambda A_i^m \approx -\sum_{\omega^n} \tilde u_{i,j}^{m - n} A_j^n,
\label{lambda_unsimplified}
\end{equation}
and for $|\omega^m| > |\lambda|$
\begin{equation}
\left(-\tau (\omega^m)^2 + \hat i \omega^m \right) A_i^m \approx \sum_{\omega^n} \tilde u_{i,j}^{m - n} A_j^n.
\label{A_m_first}
\end{equation}
Rearranging Eq. \eqref{A_m_first} gives
\begin{equation}
A_k^n  = - \sum_{\omega^l} \frac{\tau + \hat i/\omega^n}{ 1 + (\tau \omega^n)^2 } \tilde u_{k,j}^{n-l} A_j^l,
\label{A_m}
\end{equation}
for $|\omega^n| > |\lambda|$.
The first term in Eq. \eqref{lambda_unsimplified} can be neglected since $|\tau \lambda| \ll 1$ (see Eq. \eqref{lambda_assumption}). 
Therefore, substituting Eq. \eqref{A_m} in \eqref{lambda_unsimplified} yields
\begin{equation}
\lambda A_i^m  = -\sum_{|\omega^n|>|\lambda|} \sum_{\omega^l} \frac{\tau + \hat i/\omega^n}{ 1 + (\tau \omega^n)^2 } \tilde u_{i,k}^{m-n} \tilde u_{k,j}^{n-l} A_j^l + \sum_{|\omega^n|<|\lambda|} \sum_{\omega^l} \frac{1}{\lambda} \tilde u_{i,k}^{m-n} \tilde u_{k,j}^{n-l} A_j^l,
\label{lambda_A_unsimp}
\end{equation}
for $|\omega^m| < |\lambda|$.
Since the aim of this analysis is obtaining ensemble-averaged quantities, we neglect the terms with $l \ne m$ in Eq. \eqref{lambda_A_unsimp} ($\tilde u_{i,k}^{m-n}$ and $\tilde u_{i,k}^{n-l}$ are uncorrelated for $l\ne m$).
Hence
\begin{equation}
\lambda A_i^m = -\sum_{|\omega^n|>|\lambda|} \frac{\tau + \hat i/\omega^n}{ 1 + (\tau \omega^n)^2 } \tilde u_{i,k}^{m-n} \tilde u_{k,j}^{n-m} A_j^m + \sum_{|\omega^n|<|\lambda|} \frac{1}{\lambda} \tilde u_{i,k}^{m-n} \tilde u_{k,j}^{n-m} A_j^m,
\label{lambda_A_int}
\end{equation}
for $|\omega^m| < |\lambda|$.
Velocity gradients in Eq. \eqref{lambda_A_int} are a function of $|\omega^n - \omega^m|$. 
Since in the first summation $|\omega^n|>|\lambda|$ and $|\omega^m|<|\lambda|$, we neglect $\omega^m$ in this summation.
Additionally, the second summation is summed over a short range compared to the physical infinity, viz. of order $\tau_\eta^{-1}$.
Hence this term can be neglected in comparison with the first summation, which is summed over a much wider range of frequencies. 
Note that our numerical results show $|\tau_\eta \lambda| \ll 1$.
As a result Eq. \eqref{lambda_A_int} reduces to
\begin{equation}
\lambda A_i^m = -\sum_{|\omega^n|>|\lambda|} \frac{\tau + \hat i/\omega^n}{ 1 + (\tau \omega^n)^2 } \tilde u_{i,k}^{-n} \tilde u_{k,j}^{n} A_j^m,
\label{lambda_A_wo}
\end{equation}
for $|\omega^m| < |\lambda|$.
In Eq. \eqref{lambda_A_wo}, the term with $\hat i/\omega^n$ is an odd function of $\omega^n$ and can be neglected. 
Additionally, taking $\omega = \omega^n$ yields
\begin{equation}
\lambda A_i^m = -\left[ \sum_{|\omega|>|\lambda|} \frac{\tau}{ 1 + (\tau\omega)^2 } \tilde u_{i,k}^* \tilde u_{k,j} \right] A_j^m,
\label{lambda_A_l}
\end{equation}
for $|\omega^m| < |\lambda|$, in which $\tilde{\bl u}^*$ is the complex conjugate of $\tilde{\bl u}$ and is equal to $\tilde{\bl u}(-\omega)$, since $\bl u \in \mathbb R^3$.

The summation in Eq. \eqref{lambda_A_l} does not include $|\omega|<|\lambda|$.
As it was discussed earlier, a corresponding summation over these small range of frequencies is negligible compared to a summation over $|\omega|>|\lambda|$.
Therefore, we extend the summation in Eq. \eqref{lambda_A_l} to all frequencies in order to remove the dependence of the expression inside the bracket on $\lambda$.
The contribution of these small frequencies is of the order as the second summation in Eq. \eqref{lambda_A_unsimp} that was neglected earlier, hence this approximation is within the leading-order of accuracy of the final result.
This leads to
\begin{equation}
\lambda A_i^m = -\left[ \sum_{\omega} \frac{\tau}{ 1 + (\tau\omega)^2 } \tilde u_{i,k}^* \tilde u_{k,j} \right] A_j^m,
\label{lambda_A}
\end{equation}
for $|\omega^m| < |\lambda|$. 
Note that the expression inside the bracket is not a function of $\omega^m$, indicating that $\lambda$ is the same at small frequencies. 
Therefore, Eq. \eqref{lambda_A} can be simplified to Eq. \eqref{eigen_value_problem}, completing this derivation.

\end{document}